\documentclass[prc,twocolumn,superscriptaddress,showpacs,twoside,
               floatfix]{revtex4}

\usepackage{graphicx}
\usepackage{amssymb}

\begin{document}

\title{Possibility to study a two-proton halo in $^{17}$Ne.}

\author{L.\ V.\ Grigorenko}
\affiliation{Flerov Laboratory of Nuclear Reactions, JINR, RU-141980 Dubna,
Russia}

\author{Yu.\ L.\ Parfenova}
\affiliation{Sholohov st., 7, 318, RU-119634, Moscow, Russia}

\author{M.\ V.\ Zhukov}
\affiliation{Department of Physics, Chalmers University of Technology
and G\"{o}teborg University, S-41296 G\"{o}teborg, Sweden}

\date{\today. {\tt File: /latex/17ne/17ne-s-18-reresubmit.tex }}

\begin{abstract}
The nuclide $^{17}$Ne is studied theoretically in a three-body $^{15}$O+$p$+$p$ 
model. We demonstrate that the experimental condition for existence of a proton 
halo in $^{17}$Ne can be reasonably quantified in terms of $s/d$ configuration 
mixing. We discuss experimental evidences for a proton halo in $^{17}$Ne. We 
define which kind of experimental data could elucidate this issue.
\end{abstract}


%\keywords{cluster model, hyperspherical harmonic method, proton halo}

\maketitle

%###############################################################################
%\section{Introduction}
%###############################################################################

The $^{17}$Ne nucleus is an interesting and relatively poorly studied system.
It is a Borromean nucleus, since none of the binary subsystems ($^{15}$O-$p$ and
$p$-$p$) are bound. It seems to be the only realistic candidate to possess a
two-proton halo \cite{zhu95,gri03}. The level scheme was 
%reliably 
established not so long ago \cite{gui98} in multineutron transfer reactions. 
Available experimental data include Coulomb excitation \cite{chr97,chr02} and 
low energy nuclear fragmentation \cite{war98,kan03} measurements. The $^{17}$Ne 
nucleus has attracted attention also due to the possibility of two-proton
emission from the excited states \cite{chr97,gri00,gri03}. Another interesting 
issue in comparison with $^{17}$N is a $\beta$-decay asymmetry for decays to the 
first excited $1/2^+$ states in daughter nuclei \cite{bor93}.

The results of theoretical studies of $^{17}$Ne are controversial. In papers 
\cite{nak98,for01,gri03} the structure of $^{17}$Ne was studied with emphasis on 
the Coulomb displacement energy (CDE) derivation. In papers 
\cite{tim96,nak98,gup02} the $s^2$ configuration is predicted to dominate, while 
in paper \cite{for01} the dominating configuration is predicted to be
$d^2$. In paper \cite{tim96} effects of the ``halo'' kind (connected with larger 
radial extension of WF on the proton side) were considered being irrelevant for 
the $\beta$-decay asymmetry problem \cite{bor93}. However, in paper \cite{mil97} 
the $\beta$-decay asymmetry was successfully explained in these terms. It seems 
that theoretical agreement about the basic properties of $^{17}$Ne is still 
missing at the moment.

In papers \cite{kan03,jep04} the comparatively narrow core momentum distribution 
was interpreted as possible evidence for proton halo in $^{17}$Ne. This is a 
reasonable approach to the problem, as among typical experimental evidences for 
halo (e.g.\ large interaction, electromagnetic dissociation, and nucleon 
removal cross sections), the momentum distributions should give most expressed 
signal for this system. The aim of this paper is to test three-body WFs, 
obtained in \cite{gri03}, against the most recent experimental data 
\cite{chr02,kan03}. We demonstrate that the experimental question of the proton 
halo existence in $^{17}$Ne formulated as in \cite{kan03,jep04} is largely 
defined by $s/d$ configuration mixing. As we have already mentioned, the exact 
$s/d$ ratio in $^{17}$Ne is difficult to obtain unambiguously by theoretical 
calculations. To derive it from experimental data it is necessary to know the 
sensitivity of various observables to this aspect of the dynamics. We show that 
currently available experimental data are insufficient to determine reliably the 
structure (and possible halo properties) of $^{17}$Ne. We can, however, 
confidently define which kind of experimental data is required to resolve the 
puzzling issues of the $^{17}$Ne structure.

%===============================================================================

\begin{figure}
\includegraphics[width=0.42\textwidth]{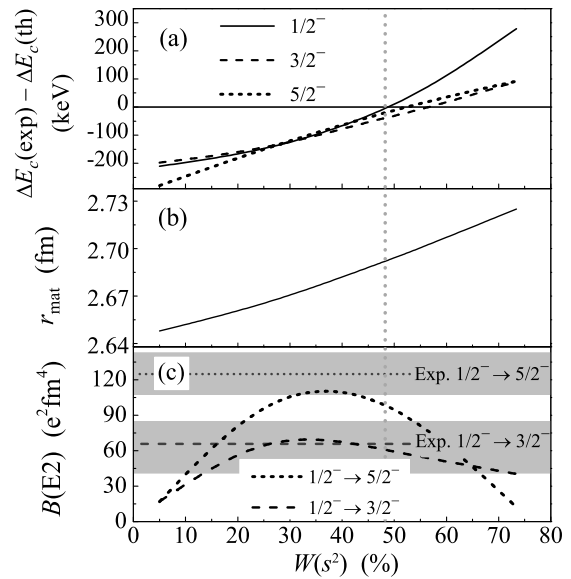}
\caption{Dependence of observables for $^{17}$Ne on the structure. (a) 
Difference of experimental and theoretical CDEs for $^{17}$N--$^{17}$Ne pair. 
(b) Matter radius. (c) $B(E2)$ probabilities for transitions between g.s.\ and 
first excited states. The vertical dotted line corresponds to $W(s^2)$ of the 
GMZ WF.}
\label{fig:dep-w-s}
\end{figure}

\textit{Structure model.}
%
%===============================================================================
%
--- Studies in this paper are based on the $^{17}$Ne WF obtained in a three-body 
model \cite{gri03}. The model predicts about  $50\,\%$ $s/d$ mixing for the 
ground state of $^{17}$Ne. Recently this 
nucleus has been studied in a three-body model \cite{gar03}, providing results 
very close to those in Ref.\ \cite{gri03}. Beside the WF from \cite{gri03}, 
which we refer to here as GMZ, we 
have also generated two WFs with high [$W(s^2)\sim 70\%$] and low [$W(s^2)\sim 
7\%$] weights of $s^2$ components. Note that this required unrealistic 
modifications of the $^{16}$F spectra. Thus, these WFs should not be regarded as 
variants of a theoretical prediction. They are used in this paper only to 
estimate a scale of the sensitivity of different observables to variations in 
$^{17}$Ne structure. Table \ref{tab:teor} and Fig.\ \ref{fig:dep-w-s} show 
various properties of the three lowest states in $^{17}$N and $^{17}$Ne 
calculated with realistic GMZ, ``high s'' and ``low s'' WFs.

Studies of the $^{17}$N--$^{17}$Ne pair as core+$N$+$N$ systems are reasonably 
well motivated. The nuclei $^{15}$N and $^{15}$O are well suited for the role of 
cores in a cluster model. Their lowest excitations are located at about 5.2 MeV 
and the lowest particle decay thresholds are at 10.2 and 7.3 MeV respectively. 
Also, in shell model studies of $^{17}$N \cite{uen96} and  $^{17}$Ne 
\cite{mil97} the admixture of excited core configurations was found to be below 
$5\,\%$, which is not enough to change ``bulk'' properties of these nuclei 
significantly. The core matter radius enters the definition of the composite 
system radius, the core charge radius is used to define a Coulomb interaction 
(if needed). For $^{15}$N the charge radius is known from electron scattering 
$r_{ch}(^{15}$N$)=2.615$ fm \cite{ajz91}. The corresponding matter radius is  
$r_{mat}(^{15}$N$)=2.49$ fm. We estimated the matter radius of $^{15}$O in two 
ways (from known experimental charge radii $r_{ch}(^{14}$N$)=2.57$ fm and 
$r_{ch}(^{16}$O$)=2.71$ fm), providing the same result: $r_{mat}(^{15}$O$)=2.53$ 
fm.

%~~~~~~~~~~~~~~~~~~~~~~~~~~~~~~~~~~~~~~~~~~~~~~~~~~~~~~~~~~~~~~~~~~~~~~~~~~~~~~~
\begin{table}
\caption{Structure and observables for $^{17}$N and $^{17}$Ne. The experimental 
CDE for $^{17}$N--$^{17}$Ne isobaric pair is $\Delta E_c(1/2^-)=7.430$ MeV. 
Properties of ground $1/2^-$ states are given in the first 6 rows. Properties of 
excited $3/2^-$ and $5/2^-$ states are given in the last 6 rows. The $B(E2)$ 
values are given in e$^2$fm$^4$. For $^{17}$N they are calculated with the rigid 
core, while those for $^{17}$Ne are corrected for experimental $B(E2)$ of 
$^{17}$N. $W(i)$ are weights of dominating WF configurations in percent.}
%
%\vspace{1mm}
%
\begin{ruledtabular}
\begin{tabular}[c]{lcccccc}
Nucleus: & \multicolumn{3}{c}{$^{17}$N} & \multicolumn{3}{c}{$^{17}$Ne} \\
WF:      & ``low s'' & GMZ  &``high s''&``low s''&  GMZ   & ``high s''    \\
\hline
 $W(s^2)$    &  7.3  &   39.8 &   63.4 &   4.8  &  48.1   &  73.4   \\
 $W(p^2)$    &  2.2  &   4.5  &   3.2  &  1.0   &   4.0   &   2.5  \\
 $W(d^2)$    & 90.4  &  55.6  &  33.0  &  94.0  &  47.8   &  23.8  \\
 $r_{mat}$ (fm)
             & 2.59  &  2.61  &  2.63  & 2.65   &  2.69   &  2.73  \\
 $\langle \rho \rangle$ (fm)
             & 4.59  &  4.81  &  5.00  & 4.82   &  5.22   &  5.49  \\
%
% $\mu$ (n.m.)& $-2.23$ & $-0.14$ & $-0.08$ & 3.13 &  0.40 & 0.35  \\
%
 $\Delta E_c$ (MeV)
             &       &        &        & 7.685  &  7.424  & 7.194  \\
 $W(sd,3/2)$ & 50.9  &  72.9  &  93.8  &  56.3  &  76.1   &  94.8  \\
 $W(d^2,3/2)$& 46.6  &  24.0  &   4.5  &  41.1  &  20.9   &   3.6  \\
 $B(E2,3/2)$ & 0.01  &  0.18  &  0.11  &  17.1  &  59.4   &  40.3  \\
 $W(sd,5/2)$ &  7.6  &  69.8  &  41.4  &   9.2  &  73.0   &  57.9  \\
 $W(d^2,5/2)$& 91.4  &  27.0  &  58.3  &  89.7  &  23.9   &  41.7  \\
 $B(E2,5/2)$ & 0.00  &  0.29  &  0.00  &  16.8  &  94.7   &  12.3  \\
\end{tabular}
\end{ruledtabular}
\label{tab:teor}
\end{table}
%~~~~~~~~~~~~~~~~~~~~~~~~~~~~~~~~~~~~~~~~~~~~~~~~~~~~~~~~~~~~~~~~~~~~~~~~~~~~~~~

%===============================================================================

\textit{CDE.} 
%
%===============================================================================
%
--- This is the only observable, for which a
sensitivity to $^{17}$Ne structure far exceeds an experimental
uncertainty. Calculations \cite{gri03} provide the WF with about
$50\,\%$ $s/d$ mixing (GMZ case) reproducing experimental CDE very
well. We rely much on this fact, as a correct CDE should guarantee
very reasonable radial characteristics of the WF. However, there
is no agreement among theorists on this issue and other checks are
also necessary.

%===============================================================================

\textit{E2 transitions.}
%
%===============================================================================
%
--- Experimental derivation of $B(E2)$ values for the first excited states of 
$^{17}$Ne is a significant advance in studies of this system: $B(E2,1/2 
\rightarrow 3/2)=66^{+18}_{-25}$ e$^2$fm$^4$ \cite{chr97} and $B(E2,1/2
\rightarrow 5/2)=124(18)$  e$^2$fm$^4$ \cite{chr02}. If we consider the $^{15}$O 
core as a rigid charged body, its contribution to $B(E2)$ of $^{17}$Ne in a 
three-body model is small due to large core mass. The $B(E2)$ values are 
underestimated by $30-50\,\%$ in such calculations. To improve the model, we 
extract $E2$ matrix element $M(E2)_{core}$ for the core from experimental
value $B(E2,1/2 \rightarrow 5/2)=6.7(1.2)$ e$^2$fm$^4$ for $^{17}$N 
\cite{til93}. It is possible, because here valence neutrons do not contribute 
the $B(E2)$ value. The resulting calculated $B(E2)$ values for different 
versions of $^{17}$Ne WFs are given in Table \ref{tab:teor} (see also Fig.\
\ref{fig:dep-w-s}). One can see that only in the case of a significant 
configuration mixing a good agreement with experimental values can be achieved.

%-------------------------------------------------------------------------------
%-------------------------------------------------------------------------------

Large, compared to ours, theoretical $B(E2)$ values were obtained in shell model 
calculations with effective charges \cite{chr02}: 105 and 155 e$^2$fm$^4$ for 
transitions to $3/2^-$ and $5/2^-$ states. Note that in our calculations there 
are no effective charges. If we recalculate our $B(E2)$ values using effective 
charges from \cite{chr02}, we get a good agreement with these calculations for
GMZ WF.

%===============================================================================

\begin{figure}[t]
\centerline{\includegraphics[width=0.48\textwidth]{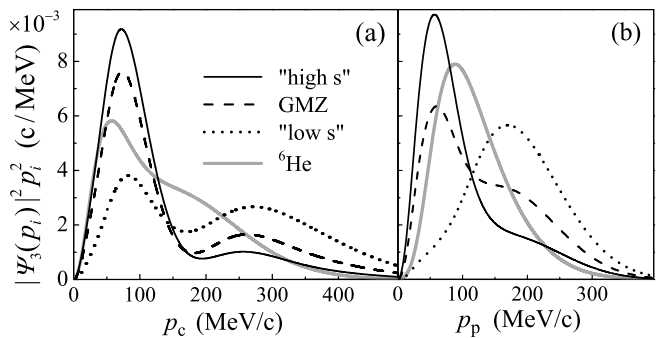}}
%\smallskip
%
\caption{Momentum distributions of a core (a) and a valence
nucleon (b) in $^6$He WF and in different $^{17}$Ne WFs.}
\label{fig:modis-pp-pcore}
\end{figure}

\textit{Momentum distributions.}
%
%===============================================================================
%
--- The first step in studies of momentum distributions from fragmentation 
reactions is to study the momentum distribution in the nucleus itself. Fig.\
\ref{fig:modis-pp-pcore} shows momentum distributions of particles from the 
valence part of the $^{17}$Ne WF. One can see that in momentum space $^{17}$Ne 
WFs have two distinctive components connected with $s^2$ and $d^2$ 
configurations. Depending on the ratio of these components momentum 
distributions for $^{17}$Ne could be either broader or narrower than the 
corresponding distributions for the $^6$He halo nucleus. For realistic GMZ WF 
the momentum distributions of constituents seem to be relatively close to that 
for $^6$He with average momenta being approximately the same.
%These distributions are more flat than in $^6$He, but the
%average momentum is approximately the same.

\begin{figure}[t]
\centerline{\includegraphics[width=0.48\textwidth]{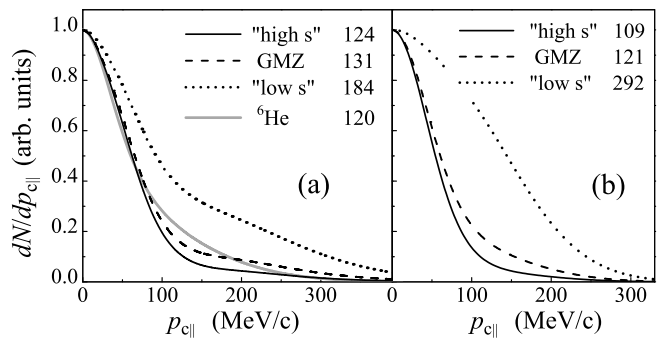}}
%\smallskip
%
\caption{Longitudinal core momentum distributions for $^6$He WF and for 
different $^{17}$Ne WFs. (a) ``No FSI'' approximation. (b) Population of four 
low-lying resonances in $^{16}$F. The values in the legends are FWHM of 
distributions in MeV/c.}
\label{fig:modis-long}
\end{figure}

The internal momentum distributions can be related to the proton sudden removal 
approximation for high energy reactions. If final state interaction (FSI) of the 
$^{15}$O core with a proton can be neglected, then measured core momentum 
distributions are simply the core momentum distributions in $^{17}$Ne (Fig.\
\ref{fig:modis-pp-pcore}a). Corresponding longitudinal momentum distributions 
(LMD) of the core are shown in Fig.\ \ref{fig:modis-long}a. Both WFs 
with large $s^2$ weights give core LMDs which are as narrow as the 
core LMD in $^6$He. Only the core distribution for ``low s'' WF has 
larger FWHM=184 MeV/c.

The inclusion of the FSI between the core and a proton (after one proton removal 
from $^{17}$Ne) can  also lead to more narrow distributions \cite{kor95,fed96}. 
Taking into account resonance states in the $^{16}$F subsystem formed after 
knockout of a valence proton the core LMD can be considered as 
\cite{fed96}
\begin{eqnarray}
%d N/dp_{c \parallel}
\frac{d N}{dp_{c \parallel}}
\, \sim  \,{\textstyle  \sum_{M}} \int d^3 p_{x} \, d^3 p_{y} \, d^2 p_{c\perp} 
\,
\delta \bigl(\textstyle \frac{15}{16} \mathbf{p}_y+\mathbf{p}_x
- \mathbf{p}_c  \bigr) \quad
\nonumber \\
 \times \; {\textstyle  \sum_{\sigma}}\left| {\textstyle  \sum_{jm}}
\langle \, \Psi^{JM}_{3} (\mathbf{X},\mathbf{Y})\, | \,\Psi ^{jm}_{2}
(\mathbf{p}_x,\mathbf{X}) \, e^{i\mathbf{p}_y\mathbf{Y}}\, \chi_p \rangle  
\right|
^{2},
\label{eq:sig-sudd}
\end{eqnarray}
where $\Psi ^{jm}_{2}(\mathbf{p}_x,\mathbf{X})$ are the WFs of $^{16}$F 
resonance states with different $j^{\pi}$, $\chi_p$ is a spin function of a 
removed proton and $\sigma$ stands for summation over spin variables. The Jacobi
coordinates $\mathbf{X}$, $\mathbf{Y}$ and the conjugated momenta 
$\mathbf{p}_x$, $\mathbf{p}_y$ are in the ``Y'' coordinate system
($\mathbf{X}$ is a distance between the core and a valence proton). This 
mechanism is dominating e.g.\ in fragmentation of $^6$He and $^{11}$Li 
\cite{aum98}. Four low-lying single-particle states in $^{16}$F are taken into 
account: $0^-$, $1^-$, $2^-$, and $3^-$ with energies 0.535, 0.728, 0.959, and
1.256 MeV above the $^{15}$O+$p$ threshold. The calculated distributions 
(\ref{eq:sig-sudd}), shown in Fig.\ \ref{fig:modis-long}b, well agree with ``no 
FSI'' approximation Fig.\ \ref{fig:modis-long}a for ``high s'' and GMZ
WFs. In the ``low s'' case the shapes of the distributions are different
(due to strong correlations in the $d^2$ WF) but the rms longitudinal 
momenta $\langle p^2_{c\parallel} \rangle^{1/2}$  for these distributions are 
reasonably close (they are 150 and 113 MeV/c for Figs.\ \ref{fig:modis-long}a 
and \ref{fig:modis-long}b respectively). It is known \cite{han96,esb96,hen96}  
that the core ``shadowing'' effect 
will lead to realistic momentum distributions which are only narrower
than those obtained in the sudden removal approximation. Thus, looking in these 
Figures one could conclude that experimental data \cite{kan03} giving FWHM
$168(17)$ MeV/c for LMD of the $^{15}$O core support case of $d^2$ 
domination in the structure of $^{17}$Ne [$W(s^2)<25\,\%$].
There is, however, an obstacle which makes the analysis of the situation more 
complicated.

%-------------------------------------------------------------------------------
%-------------------------------------------------------------------------------

%===============================================================================

\textit{Interaction cross sections.}
%
%===============================================================================
%
--- Interaction and proton removal cross sections are calculated in the eikonal 
approximation of the Glauber model \cite{ber98} for three-body $^{17}$Ne 
nucleus. In this model breakup cross sections are related to interaction cross 
sections of the fragments as
\begin{equation}
\sigma _{str}^{1p}+\sigma _{str}^{2p}+\sigma _{dif} =\sigma _{-2p}=\sigma
_{I}(^{17}\text{Ne})-\sigma _{I}(^{15}\text{O})\;.
\end{equation}
In our calculations the cross sections are determined by the interaction
potential \cite{hen96} generated from the free $NN$-interaction  \cite{ray79}
and nuclear fragment densities.

%-------------------------------------------------------------------------------
%-------------------------------------------------------------------------------

The $^{9}$Be density $\rho $ is parameterized by the modified harmonic 
oscillator expression \cite{vri87} with $a$=1.791 fm
\begin{equation}
\rho (r)=\rho _{0}[1+\alpha (r/a)^{2}]\exp [-(r/a)^{2}]\,, 
\label{dens}
\end{equation}
which gives the $^{9}$Be charge radius 2.52 fm.
%(parameter $\alpha $ is related to $a$ \cite{vri87}).
The $^{12}$C and $^{28}$Si densities are approximated by the sum of Gaussians
%
%\[
%\rho (r)=\textstyle\sum_{i}A_{i}\left( e^{- (r-R_{i})^2/\gamma^2}
%+e^{- (r+R_{i})^2/\gamma^2}\right) \,.
%\]
%
with parameters from Ref.\ \cite{vri87}.

The $^{15}$O density distribution is not known; we approximate it
by the two-parameter Fermi expression \cite{vri87}
\begin{equation}
\rho (r)=\rho _{0}/\left(1+\exp [(r-c)/z]\,\right)\,. \label{GausDens}
\end{equation}
Parameters $c = 3.266$ fm and $z=0.1$ fm are chosen to reproduce both the 
$^{15}$O matter radius and interaction cross sections for reactions 
$^{15}$O+$^{28}$Si at energies $22-44$ MeV/amu \cite{war98} (Table 
\ref{tab:o-si}) and $^{15}$O+$^{9}$Be, $^{15}$O+$^{12}$C at the energy 710 
MeV/amu \cite{oza01} (Table \ref{tab:be-high}).

The above choice of core and target densities allows us to reproduce the 
experimental data on the $p$+$^{28}$Si \cite{bar54} and $^{17}$Ne+$^{28}$Si 
\cite{war98} interaction cross sections at energies $20-50$ MeV/amu \cite{war98} 
(Table \ref{tab:ne-si}). The agreement with experiment for p, $^{15}$O, 
$^{17}$Ne interaction cross sections on $^{9}$Be and $^{12}$C targets is also 
very good for two available experimental energies (Table \ref{tab:be-high}). All 
results for $^{17}$Ne in Tables \ref{tab:be-high} and \ref{tab:ne-si}
are calculated with the GMZ WF.
%
%All calculated interaction cross sections for different model $^{17}$Ne WFs are 
%consistent with available experimental values. Thus this charecteristic 
%is not sufficiently sensitive to draw some conclusions about structure. The 
%matter radii for our WFs (Fig.\ \ref{fig:dep-w-s}) are also in reasonable 
%agreement with $r_{mat}^{exp}=2.75(7)$ fm extracted in \cite{oza94} 
%using  Glauber approach with harmonic-oscillator densities.  
%
%All calculated $^{17}$Ne interaction cross sections are consistent with 
%available experimental values. 
The matter radius for our WF (Table \ref{tab:teor}) is also in an agreement with 
effective $r_{mat}^{exp}=2.75(7)$ fm extracted in  \cite{oza94} using  Glauber 
model with harmonic-oscillator densities.

%===============================================================================
\begin{table}
\caption{Experimental \cite{war98} and theoretical interaction cross sections
$\sigma_I$  for the $^{15}$O+$^{28}$Si reaction (in mb).}
%
%\vspace{2mm}
\begin{ruledtabular}
\begin{tabular}{cccc}
$E_{\text{beam}}$ (MeV/amu) & 22.0$-$30.8 & 30.8$-$38.0 & 38.0$-$44.0  \\
\hline
 $\sigma _{I}$(exp)    & 1740(40)     & 1790(40)     &  1680(40)  \\
 $\sigma_{I}$(th)      & 1860         & 1780         &  1725          \\
\end{tabular}
\end{ruledtabular}
\label{tab:o-si}
\end{table}
%===============================================================================

%===============================================================================
\begin{table}
\caption{Experimental and theoretical cross sections (in mb) for $p$, $^{15}$O, 
and $^{17}$Ne on different targets at 710 and 66 MeV/amu. The experimental 
values for $^{17}$Ne from \cite{oza01}, measured at the energy 680 MeV/amu, are 
scaled according to the energy dependence of the interaction cross section.}
%
%\vspace{2mm}
%
\begin{ruledtabular}
\begin{tabular}{ccccc}
target & $\sigma_{I}$($p$) & $\sigma _{I}(^{15}$O) & $\sigma_{I}(^{17}$Ne) &
  $\sigma _{-2p}(^{17}$Ne)   \\
\hline
\multicolumn{4}{c}{$E_{\text{beam}}=710$ MeV/amu}  \\
Be(exp) & $214(13)$ \cite{bar54} & 912(23) \cite{oza01} & $972(45)$
\cite{oza01} & \\
Be(th)  & 210                      & 914                      & 987
   &
73  \\
C(exp) & $232(14)$ \cite{bar54} & 922(49) \cite{oza01} & $1094(76)$
\cite{oza01} & \\
C(th)  & 240                    & 970                      & 1050
 &
80  \\
\multicolumn{4}{c}{$E_{\text{beam}}=66$ MeV/amu}  \\
Be(exp)         & 316 \cite{bar54} &      &      & 191(48) \cite{kan03} \\
Be(th)          & 308              & 1070 & 1179 & 109                      \\
\end{tabular}
\end{ruledtabular}
\label{tab:be-high}
\end{table}
%===============================================================================

%===============================================================================
\begin{table}
\caption{Experimental \cite{war98} and theoretical interaction and $2p$ removal
cross sections for the $^{17}$Ne+$^{28}$Si reaction (in mb).}
%
%\vspace{2mm}
\begin{ruledtabular}
\begin{tabular}{ccccccc}
$E_{\text{beam}}$ (MeV/amu) & $\sigma _{I}$(exp) & $\sigma
_{-2p}$(exp)  &
$\sigma _{I} $(th) & $\sigma _{-2p} $(th)\\
\hline
 27.6$-$37.7 & 1980(70) &              & 1950 & 155 \\
 37.7$-$46.3 & 1930(70) &              & 1868 & 149 \\
 46.3$-$53.3 & 1770(70) &              & 1813 & 145 \\
 46.         &              & 260(30)  &      & 147 \\
\end{tabular}
\end{ruledtabular}
\label{tab:ne-si}
\end{table}
%===============================================================================

%===============================================================================

\textit{Proton removal from halo in $^{17}$Ne.}
%
%===============================================================================
%
--- Contrary to the total interaction cross sections, the $2p$ removal cross 
sections are $30-40\%$ underestimated in our calculations (see Tables
\ref{tab:be-high} and \ref{tab:ne-si}). To check the sensitivity of the cross 
sections to variations of the $^{17}$Ne structure we have calculated the $2p$ 
removal cross sections for $^{17}$Ne on Be target at 66 MeV/amu with different 
$^{17}$Ne WFs. The corresponding $\sigma_{-2p}$ are $120$ mb, $109$ mb and $82$ 
mb for ``high s'', GMZ and ``low s'' WFs. These results show that this variation 
of the $^{17}$Ne structure is not sufficient to compensate for the discrepancy 
with experiment.

To overcome this problem, it
%the theoretical part of
was suggested in Ref.\ \cite{kan03} that (i) the halo is very large ($\langle 
r_p \rangle \sim 4.5$  and 3.8 fm for pure $s^2$ and $d^2$ configurations
compared to $\langle r_p \rangle \sim 3.7$, 3.5 and 3.3 fm given by ``high s'', 
GMZ and ``low s'' WFs) and (ii) the matter radius of the $^{15}$O core is small 
($r_{mat}=2.42$ fm compared to 2.53 fm in this work). Only these (too strong, in 
our opinion) assumptions provided $\sigma_{-2p}($th$) \sim 168$ mb \cite{kan03}
for pure $s^2$ configuration in an 
%qualitative 
agreement with experiment. In our model, the halo size is fixed by CDE and a
reduction of the core size leads to a deterioration of the agreement for 
multiple calculated reaction cross sections. We do not feel there is a freedom 
in that direction and other explanations are required.

%-------------------------------------------------------------------------------
\begin{figure}
\includegraphics[width=0.47\textwidth]{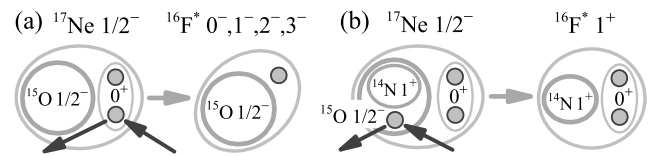}  
\caption{Dominating reaction mechanisms for one-proton knockout from $^{17}$Ne.
(a) $s$/$d$-wave proton knockout from halo, populating negative parity states
in $^{16}$F. (b) $p$-wave proton knockout from $^{15}$O core, populating $1^+$
states in $^{16}$F.}
\label{mechs}
\end{figure}
%-------------------------------------------------------------------------------

The calculated $\sigma_{-2p}$ values (see Table \ref{tab:be-high}) for 710 
MeV/amu on C target  of about $40$ mb ``per proton'' (for proton removal from 
the halo) are in a qualitative agreement with theoretical proton knockout cross 
section from $^8$B (about 80 mb \cite{par02}) which are also in a good agreement 
with experimental data. It is expected that in $^8$B the halo feature is more
expressed than in $^{17}$Ne due to smaller Coulomb interaction and smaller 
binding energy. Also, the $^7$Be core in $^8$B is smaller than the $^{15}$O core 
in $^{17}$Ne increasing the probability of the $^7$Be core survival. Otherwise, 
if we explain the whole two-proton removal cross section in $^{17}$Ne 
\cite{kan03} as a removal from halo we come to a contradiction. From this cross
section it should then be concluded that in $^{17}$Ne the halo is much more 
pronounced than in $^8$B (which is not in accord with general expectations) 
whereas from momentum distribution \cite{kan03} (which is relatively broad) a 
pronounced halo in $^{17}$Ne should not be expected.

%-------------------------------------------------------------------------------
\begin{figure}[t]
\centerline{\includegraphics[width=0.48\textwidth]{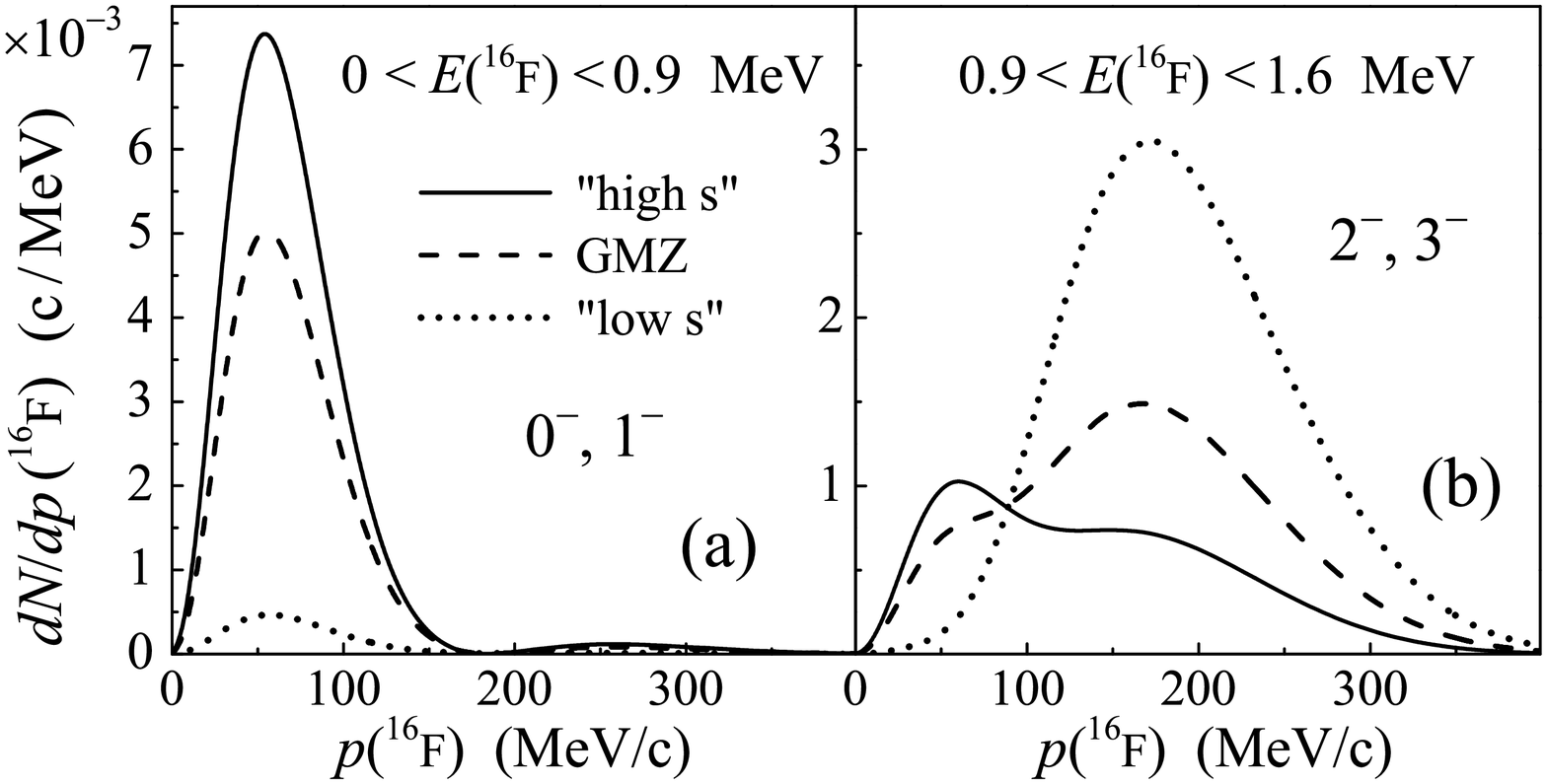}}
%\smallskip
\caption{Momentum distribution of $^{16}$F cm for proton knockout from
$^{17}$Ne gated on the energy ranges with $s$-wave (a) and $d$-wave (b) negative 
parity states in $^{16}$F.}
\label{fig:modis-invar}
\end{figure}
%-------------------------------------------------------------------------------

%===============================================================================

\textit{Proton removal from the $^{15}$O core.}
%
%===============================================================================
%
--- The possible solution of the above problem is incorporates  processes which 
are beyond a simple valence nucleon removal. For the case of $^{11}$Be (one
neutron halo nucleus) it was shown in papers \cite{tos99,aum00,par00} that 
beside the valence nucleon removal, the removal of a tightly bound core nucleon 
leading to low-lying excited states of a fragment can also give an important
contribution to the cross section. For the $^{17}$Ne case it means that a 
process of $p$-wave proton removal from $^{15}$O core has to be considered (see 
also \cite{kan03}, ``model-3''). The simplest possible mechanism is 
schematically illustrated in Fig.\ \ref{mechs}. A $p$-wave proton knockout from 
the $^{15}$O ($1/2^-$) core leads to $^{14}$N in $1^+$ states. These states
together with the valence protons (which are predominantly in the $0^+$ relative 
motion state in $^{17}$Ne) could populate $1^+$ states in $^{16}$F located below 
the $^{14}$N+$2p$ threshold. These states decay only via $^{15}$O+$p$ channel 
and thus contribute the two-proton removal cross section for $^{17}$Ne.

The calculated cross section of the $p_{1/2}$ proton removal from the $^{15}$O 
nucleus with the proton separation energy $S_p=7.279$ MeV is $\sigma _{-p}= 
19.4$ mb and the FWHM of the LMD is 177 MeV/c. The removal cross section of the 
 $p_{3/2}$ proton with $S_p=11.247$ MeV is $\sigma _{-p}$=15.3 mb and FWHM=200 
MeV/c. Taking into account two protons in the $p_{1/2}$ state and four protons 
in the $p_{3/2}$ state, we get an assessment of the proton removal cross section 
100 mb and FWHM=190 MeV/c, that is in a good agreement with the 
experimental data $80(10)$ mb and $190(10)$ MeV/c from \cite{jep04} for the beam 
energy 56 MeV/amu. The cross section of the proton removal from the $^{15}$O 
core is obtained in the three-body model similarly to \cite{par00}:
$\sigma _{-p}=53$ mb. Together with the $2p$ removal from halo (Table 
\ref{tab:be-high}) this provides the total $2p$ removal cross section of 162 mb, 
which is in an agreement with the results from \cite{kan03}. Thus, broad 
momentum distribution [$168(17)$ MeV/c] found in \cite{kan03} can not be a proof 
of $d^{2}$ domination in the $^{17}$Ne halo as these data are presumably 
strongly influenced by the processes on the core.

%===============================================================================

\textit{Invariant mass measurement of $^{16}$F.}
%
%===============================================================================
%
--- It is easy to disentangle halo and core contributions to the two-proton 
removal cross section in an exclusive experiment. The invariant mass measurement 
of $^{15}$O and proton should allow to distinguish the processes of proton knock 
out from halo (which should mainly proceed through low-lying negative parity 
states in $^{16}$F) and proton knock out from the core (which involves $1^+$ 
states of $^{16}$F). From simple spectroscopic considerations the populations of 
the energy ranges for relative motion of $p$ and $^{15}$O corresponding to 
$0^-$, $1^-$, $2^-$, and $3^-$ states in $^{16}$F are proportional to  
$\frac{1}{4}W(s^2)$,  $\frac{3}{4}W(s^2)$,  $\frac{1}{4}W(d^2)$, and  
$\frac{7}{20}W(d^2)$ in the first approximation. Real situation could be more 
complicated and exclusive momentum distributions can help to improve the 
understanding. The momentum distributions of $^{16}$F cm calculated in the same 
model as Eq.\ (\ref{eq:sig-sudd}) and gated on different ranges of excitation 
energy in $^{16}$F (where there are only negative parity states) are shown in 
Fig.\ \ref{fig:modis-invar}. If the reaction mechanism of the model Eq.\ 
(\ref{eq:sig-sudd}) prevails, such experimental distributions should be free 
from the core contributions. Moreover, the ratii and shapes of the corresponding
distributions in Figs.\ \ref{fig:modis-invar}a and \ref{fig:modis-invar}b are 
strongly sensitive to the structure of halo in $^{17}$Ne. So, comparison of such 
distributions could make it possible to obtain conclusive information on this 
issue.

%===============================================================================
%
\textit{Conclusion.}
%
%===============================================================================
%
--- The question of an  existence of a proton halo in $^{17}$Ne, as it is 
approached from experimental side, can be quantified as the question of $s/d$ 
configuration 
mixing. In case of significant (say, $\geq 50 \%$) $s$-wave component in the 
$^{17}$Ne WF the ``classical'' fingerprints of the halo should exist, e.g.\ 
narrow core momentum distributions for valence proton knock out. These 
distributions 
should have comparable width to the corresponding distributions in $^6$He case, 
which is a recognized example of halo nucleus. There is considerable 
experimental evidence [CDE, $B($E2)] that the halo part of $^{17}$Ne WF is a 
significant mixture of $s^2$ and $d^2$ configurations.

The proton removal from halo is likely to be responsible only for $60 - 70\, \%$ 
of the two-proton removal cross section from $^{17}$Ne. The rest is possibly 
connected with the proton removal from the core. Thus consideration of inclusive 
LMD of the core is insufficient to draw conclusions about the halo 
property of $^{17}$Ne as this characteristic possibly has large contribution 
from processes on core. The question about configuration mixing in $^{17}$Ne can 
be resolved by invariant mass measurement of $^{15}$O and $p$ after proton 
knockout. 
%Gating momentum distribution of $^{16}$F on the low-energy excitations 
%of $^{15}$O+$p$ the distributions are obtained which are highly sensitive to 
%the structure of $^{17}$Ne.

%===============================================================================
%
\textit{Acknowledgments.}
%
%===============================================================================
%
--- The authors acknowledge the financial support from the Royal Swedish Academy 
of Science, Russian RFBR Grant 00-15-96590 and Russian Ministry of Industry and 
Science grant NS-1885.2003.2.

%###############################################################################

%###############################################################################

\end{document}